\documentclass[12pt]{article}

\usepackage[margin=1.3in]{geometry}

\usepackage{cite}

\usepackage{amssymb,latexsym}

\usepackage{epstopdf}

\usepackage{graphics,epsfig}

\usepackage{color}
\usepackage{xcolor}

\usepackage{amsmath,amsfonts}

\usepackage{mathrsfs}

\usepackage{upgreek}

\DeclareMathAlphabet{\mathpzc}{OT1}{pzc}{m}{it}

\usepackage{feynmp}
\usepackage{feynmp-auto}

\newcommand*{\Scale}[2][4]{\scalebox{#1}{$#2$}}%

\let\a=\alpha \let\b=\beta \let\g=\gamma \let\d=\delta \let\e=\epsilon
\let\z=\zeta  \let\th=\theta  \let\k=\kappa
\let\l=\lambda \let\m=\mu \let\n=\nu \let\x=\xi \let\p=\pi %\let\r=\rho
\let\s=\sigma \let\t=\tau  \let\f=\phi  
 
\let\w=\omega      \let\G=\Gamma  \let\Th=\Theta \let\L=\Lambda
\let\X=\Xi  \let\S=\Sigma  \let\Y=\Psi
 
\let\la=\label  
  
\def\nn{\nonumber} \def\bd{\begin{document}} \def\ed{\end{document}}
\def\ds{\documentstyle} \let\fr=\frac \let\bl=\bigl \let\br=\bigr
\let\Br=\Bigr \let\Bl=\Bigl
\let\bm=\bibitem
\let\na=\nabla
\def\tU{{\widetilde U}}
\let\pa=\partial \let\ov=\overline
\def\ie{{\it i.e.\ }}
\newcommand{\be}{\begin{equation}}
\newcommand{\ee}{\end{equation}}
\def\ba{\begin{array}}
\def\ea{\end{array}}
\def\ft#1#2{{\textstyle{{\scriptstyle #1}\over {\scriptstyle #2}}}}
\def\fft#1#2{{#1 \over #2}}
\def\F#1#2{{ F_{#1}^{(#2)} }}
\def\cF#1#2{{ {\cal F}_{#1}^{(#2)} }}

\def\R{{\bf R}}
\def\sst#1{{\scriptscriptstyle #1}}
\def\oneone{\rlap 1\mkern4mu{\rm l}}
\def\e7{E_{7(+7)}}
\def\td{\tilde}
\def\wtd{\widetilde}
\def\im{{\rm i}}
\def\bog{Bogomol'nyi\ }
\newcommand{\ho}[1]{$\, ^{#1}$}
\newcommand{\hoch}[1]{$\, ^{#1}$}
\newcommand{\bea}{\begin{eqnarray}}
\newcommand{\eea}{\end{eqnarray}}
\newcommand{\ra}{\rightarrow}
\newcommand{\lra}{\longrightarrow}
\newcommand{\Lra}{\Leftrightarrow}
\newcommand{\ap}{\alpha^\prime}
\newcommand{\bp}{\tilde \beta^\prime}
\newcommand{\cB}{{\cal B}}
\newcommand{\cO}{{\cal O}}
\newcommand{\vecx}{\vec{x}}
\newcommand{\vecy}{\vec{y}}
\newcommand{\vecp}{\vec{p}}
\newcommand{\vecq}{\vec{q}}
\newcommand{\tr}{{\rm tr} }
\newcommand{\Tr}{{\rm Tr} }
\newcommand{\NP}{Nucl. Phys. }

\newcommand{\cL}{{\cal L}}
\newcommand{\cA}{{\cal A}}
\newcommand{\cT}{{\cal T}}
\newcommand{\cR}{{\cal R}}
\newcommand{\cD}{{\cal D}}
\newcommand{\cH}{{\cal H}}

\def\Cb{\bar{C}}

\def\sst#1{{\scriptscriptstyle #1}}
\def\0{{\sst{(0)}}}
\def\1{{\sst{(1)}}}
\def\2{{\sst{(2)}}}
\def\3{{\sst{(3)}}}
\def\4{{\sst{(4)}}}
\def\5{{\sst{(5)}}}
\def\6{{\sst{(6)}}}
\def\7{{\sst{(7)}}}
\def\8{{\sst{(8)}}}
\def\9{{\sst{(9)}}}
\def\p{{\sst{(p)}}}
\def\q{{\sst{(q)}}}
\def\ve{\varepsilon}
\def\vf{\varphi}
\def\F{\Phi}
\def\wg{\wedge}

\def\thb{\bar{\theta}}
\def\Thb{\bar{\Theta}}
\def\barp{\bar{p}}
\def\barq{\bar{q}}
\def\barc{\bar{c}}
\def\bard{\bar{d}}
\def\e{\epsilon}

\def \bi{\bibitem}
\def \la {\label}

\def \l {\lambda}
\def\foot{\footnote}
\def \tl  {{\tilde \l}}
\def \sql {{\sqrt \l}}
\def \adss {$AdS_5 \times S^5$\ }
\newcommand{\rf}[1]{(\ref{#1})}
\def \ov {\over}

\def\th{\theta}
\def\Th{\Theta}
\def\vth{\vartheta}
\def\btheta{{\bar\theta}}
\def\ttheta{{{\tilde\theta}}}
\def\bttheta{{{\bar\ttheta}}}
\def\vth{\vartheta}

\def\ra{\rightarrow}
\def\N{\nabla}
\def\F{{\cal F}}
\def\uM{\underline{M}}
\def\uA{\underline{A}}
\def\uN{\underline{N}}
\def\uP{\underline{P}}
\def\ua{\underline{a}}
\def\ub{\underline{b}}
\def\uc{\underline{c}}
\def\ud{\underline{d}}
\def\ue{\underline{e}}
\def\uf{\underline{f}}
\def\ui{\underline{i}}
\def\uj{\underline{j}}
\def\uk{\underline{k}}
\def\ul{\underline{l}}
\def\ual{\underline{\alpha}}
\def\ube{\underline{\beta}}
\def\um{\underline{m}}
\def\un{\underline{n}}
\def\up{\underline{p}}
\def\uq{\underline{q}}
\def\ur{\underline{r}}
\def\us{\underline{s}}
\def\umu{\underline{\mu}}
\def\unu{\underline{\nu}}
\def\ula{\underline{\l}}
\def\uka{\underline{\k}}
\def\usi{\underline{\s}}
\def\urh{\underline{\r}}
\def\cc{\circ}
\def\eqv{\equiv}

\def\ni{\noindent}

\def\Ep{E^{{}^{(+)}}}
\def\Em{E^{{}^{(-)}}}

\def\Mp{M^{{}^{(+)}}}
\def\Mm{M^{{}^{(-)}}}

\def \ha{{1\ov 2}}

\def\r{\rho}

\def\Y{{\rm Y}}
\def\X{{\rm X}}
\def\tY{\tilde{\rm Y}}
\def\tX{\tilde{\rm X}}
\def\dY{\dot{\rm Y}}
\def\dX{\dot{\rm X}}

\def \J {\mathcal{J}}
\def \del {\partial}

\def\dF{\dot{F}}
\def\dG{\dot{G}}
\def\df{\dot{f}}
\def \E {{\cal E}}
\def \S {{\cal S}}
\def \J {{\cal J}}

\def\ms{\mathcal{S}}
\def\mj{\mathcal{J}}
\def\soj{\fr{\ms}{\mj}}
\def \R {{\bf R}}
\def \om {\omega}
\def \bE {\bar E}
\def \x {{\cal X}}

\def \bi{\bibitem}
\def \la {\label}

\def \l {\lambda}
\def\foot{\footnote}
\def \tl  {{\tilde \l}}
\def \sql {{\sqrt \l}}
\def \adss {$AdS_5 \times S^5$\ }
\def \ov {\over}

\def \varpi {{\rm w}}

\def\thb{\bar{\theta}}
\def\Thb{\bar{\Theta}}
\def\mb{\bar{\m}}
\def\ab{\bar{\a}}
\def\zb{\bar{z}}
\def\psib{\bar{\psi}}
\def\barp{\bar{p}}
\def\barq{\bar{q}}
\def\barc{\bar{c}}
\def\bard{\bar{d}}
\def\e{\epsilon}
\def\wb{\bar{w}}
\def\lb{\bar{\l}}
\def\Jb{\bar{J}}
\def\Nb{\bar{N}}
\def\Zb{\bar{Z}}
\def\pab{\bar{\pa}}

\def\At{\tilde{A}}
\def\Bt{\tilde{B}}
\def\Ct{\tilde{C}}
\def\Dt{\tilde{D}}
\def\Et{\tilde{E}}
\def\Ft{\tilde{F}}
\def\Gt{\tilde{G}}
\def\Ht{\tilde{H}}
\def\Kt{\tilde{K}}
\def\Mt{\tilde{M}}
\def\Nt{\tilde{N}}
\def\Rt{\tilde{R}}
\def\at{\tilde{a}}
\def\bt{\tilde{b}}
\def\ct{\tilde{c}}
\def\dt{\tilde{d}}
\def\et{\tilde{e}}
\def\ft{\tilde{f}}
\def \ztt{\tilde{\z}}
\def \zetat{\tilde{\zeta}}
\def\htil{\tilde{h}}
\def\gt{\tilde{g}}
\def\nt{\tilde{n}}
\def\mut{\tilde{\mu}}
\def\nut{\tilde{\nu}}
\def\pht{\tilde{\f}}
\def\Phit{\tilde{\Phi}}
\def\vft{\tilde{\vf}}

\def\rht{\tilde{\rho}}

\def\asth{\hat{*}}
\def\phh{\hat{\phi}}

\def\bA{{\bf A}}

\def\ola{\overleftarrow}
\def\ora{\overrightarrow}
\def\alt{\tilde{\a}}

\def\eh{\hat{e}}
\def\eph{\hat{\e}}
\def\ph{\hat{p}}
\def\alh{\hat{\a}}
\def\beh{\hat{\b}}
\def\gah{\hat{\g}}
\def\Fh{\hat{F}}
\def\muh{\hat{\m}}
\def\nuh{\hat{\n}}
\def\thh{\hat{\th}}
\def\rhh{\hat{\r}}
\def\dh{\hat{d}}
\def\ih{\hat{i}}
\def\jh{\hat{j}}
\def\hh{\hat{h}}
\def\nh{\hat{n}}
\def\gh{\hat{g}}
\def\kh{\hat{k}}
\def\deh{\hat{\d}}
\def\wh{\hat{w}}
\def\lah{\hat{\l}}
\def\Ah{\hat{A}}
\def\Gh{\hat{G}}
\def\Kh{\hat{K}}
\def\Nh{\hat{N}}
\def\Rh{\hat{R}}
\def\Ch{\hat{C}}
\def\Omh{\hat{\Omega}}

\def\xh{\hat{x}}

\def\ps{\rlap{\, /}\;\,p }
\def\ks{\rlap{\, /}\;\,k }

\def\gym{g_{YM}}

\def\adot{\dot{a}}
\def\bdot{\dot{b}}
\def\bpa{\bar{\pa}}

\def\pr{\prime}
\def\ssk{\medskip}
\def\clb{\color{blue}}
\def\clr{\color{red}}
\def\clg{\color{green}}
\def\clp{\color{purple}}
\def\clc{\color{cyan}}
\def\clm{\color{magenta}}
\def\cly{\color{yellow}}

\def\bfA{{\bf A}}
\def\bfB{{\bf B}}
\def\bfK{{\bf K}}
\def\bfU{{\bf U}}
\def\bfX{{\bf X}}
\def\bfY{{\bf Y}}
\def\bfZ{{\bf Z}}
\def\bfg{{\bf g}}
\def\bfn{{\bf n}}

\def\bsk{\bigskip}
\def\ssk{\medskip}

\def\Ec{{\cal E}}

\begin{document}

\overfullrule=0pt
\parskip=2pt
\parindent=12pt
\headheight=0in \headsep=0in \topmargin=0in
\oddsidemargin=0in

\vspace{ -3cm}
\thispagestyle{empty}
%\vspace{1cm}
%\begin{flushright}
%Preprint DFPD 01/TH/\\
%hep-th/
%\end{flushright}

 \vspace{0.1cm}

\setcounter{equation}{0}
\setcounter{footnote}{0}
\setcounter{section}{0}

\begin{center}

{\Large\bf  Quantization of gravity and finite temperature effects}

\vskip 0.8cm

%
%\vspace{0.5cm}
%
%A. J. Nurmagambetov$\,^{\spadesuit}$\let\thefootnote\relax\footnotetext{$^{\spadesuit}$ Also at {\it Karazin Kharkov National University, 4 Svobody Sq., Kharkov, UA 61022} \& {\it Usikov Institute for Radiophysics and Electronics, 12 Proskura St., Kharkov, UA 61085}. }
I. Y. Park
\\
%
%\vspace{0.3cm}
%
%$^{\spadesuit}$
%{\it Akhiezer Institute for Theoretical Physics of
%NSC KIPT,\\
%1 Akademicheskaya St., Kharkov, \\ UA 61108 Ukraine \\
%ajn@kipt.kharkov.ua
%}
%
%\vspace{0.3cm}
{\it Department of Applied Mathematics,
Philander Smith College %\footnote{Home institute}
                               \\
Little Rock, AR 72223, USA \\
inyongpark05@gmail.com
}

 \vspace{.5cm}

\end{center}

 \vspace{0.1cm}

\begin{abstract}

Gravity is perturbatively renormalizable for the physical states which can be conveniently defined via foliation-based quantization. In recent sequels, one-loop analysis was explicitly carried out for Einstein-scalar and Einstein-Maxwell systems. Various germane issues and all-loop renormalizability have been addressed. In the present work we make further progress by carrying out several additional tasks.  Firstly, we present an alternative 4D-covariant derivation of the physical state condition by examining gauge choice-independence of a scattering amplitude. To this end, a careful dichotomy between the ordinary, and large gauge symmetries is required and appropriate gauge-fixing of the ordinary symmetry must be performed. Secondly, vacuum energy is analyzed in a finite-temperature setup. A variant optimal perturbation theory is implemented to two-loop. The renormalized mass determined by the optimal perturbation theory turns out to be on the order of the temperature, allowing one to avoid the cosmological constant problem. The third task that we take up is examination of the possibility of asymptotic freedom in finite-temperature quantum electrodynamics. In spite of the debates in the literature, the idea remains reasonable.

\end{abstract}
\newpage

%%%%%%%%%%%%%%%%%

%\vspace{.3in}

%\ni {\bf Acknowledgments}

%\ni The research of this work was funded in part by Hangyang University, South Korea.

%%%%%%%%%%%%%%%%%%%%%%%%%%%%%%
%%%%%%%%%%%%%%%%%%%%%%%%%%%%%%
\section{Introduction}
%%%%%%%%%%%%%%%%%%%%%%%%%%%%%%
%%%%%%%%%%%%%%%%%%%%%%%%%%%%%%

Although quantum gravitational effects (reviews of various approaches to quantization of a matter-gravity system can be found, e.g., in \cite{Thiemann:2007zz,Buchbinder,Birrell,Esposito,Parker}) are often set aside, it has become increasingly clear through a series of recent works \cite{Park:2019amz,Park:2017wiw,Nurmagambetov:2020ann,Park:2017dib,Park:2021ohu,Park:2018xtt,Nurmagambetov:2018het,Nurmagambetov:2019mih,Nurmagambetov:2019bqz,Park:2019lbj,Kawai:2017txu,Kawai:2020rmt,Ho:2020cbf,Ho:2020cvn} that they are crucial for solving some of the outstanding problems in theoretical and astro-theoretical physics. For instance, they have proven to be indispensable for precisely formulating (and potentially solving) the black hole information paradox \cite{Park:2013rm}\cite{Park:2017wiw,Park:2018xtt,Park:2019lbj}.\footnote{The quantum-gravitational account of the black hole information put forth in \cite{Park:2017wiw} and \cite{Park:2019lbj} is that the system information evolves unitarily, where a non-perturbative bounce solution (see, e.g., \cite{Christodoulou:2016vny,Bianchi:2018mml,BenAchour:2020mgu,BenAchour:2020gon} for earlier bounce solutions and their roles in black hole information) plays an important role in entanglement among the system components.} Quantized gravity ought also to be an optimal arena for a systematic formulation (and resolution) of cosmological constant problem (see, e.g., \cite{Padmanabhan:2002ji,Martin:2012bt,Sola:2013gha} for reviews of the problem), since the resolution would require renormalization of the vacuum energy. A systematic analysis of vacuum energy has recently been conducted in a quantized gravity setup with finite-temperature \cite{Park:2021ohu}. In the present work we provide a brief review of some of these developments, and press on.

For quantization, there are many aspects of the analysis with which one must be concerned. These include boundary conditions \cite{Regge:1974zd}\cite{Esposito}\cite{Park:1998yw}\cite{Park:2018xtt}, identification of the physical states \cite{Higuchi:1991tk}\cite{Gay-Balmaz:2010anx}\cite{Park:2015ybl}, removal of the trace mode of the fluctuation metric \cite{Ortin}\cite{Park:2014tia}, technical but crucial issues surrounding the background field method (BFM) \cite{Kallosh:1978wt}\cite{Antoniadis:1995fc}\cite{Park:2014noa,Park:2015ota,Park:2016zgt,Park:2018vci}, and gauge choice-(in)dependence \cite{Vilkovisky:1984st,Fradkin:1983nw,Modesto:2017hzl,Huggins:1987zw,Odintsov:1991fk,Falls:2015qga}\cite{Park:2018vci}. For all of these it is crucial to carefully analyze the gauge symmetries, including large gauge transformations (LGTs) \cite{Harvey:1996ur}\cite{DiVecchia:1998ky}. The presence of large gauge symmetry makes the subject complicated but, at the same time, rich. For one thing it clearly demonstrates the necessity of  Hilbert space extension by including non-Dirichlet boundary conditions. Our previous works recognizing the importance of the non-Dirichlet boundary conditions include \cite{Park:2013vpa,Park:2015xoa,James:2016zha}. They were subsequently analyzed in \cite{Park:2018xtt}\cite{Park:2018vci}. Derivation of the physical states and some pertinent analyses can be found in the early sequels of \cite{Park:2014noa}\cite{Park:2015ybl}. The pathology associated with the trace mode  \cite{Gibbons:1978ac}\cite{Mazur:1989by} was reviewed in \cite{Park:2014tia,Park:2015ota,Park:2016zgt} and its removal by gauging away was presented in \cite{Park:2015ota,Park:2015xoa,Park:2016zgt,Park:2018vci}. A refined application of the BFM \cite{Park:2015ota}\cite{Park:2016zgt}\cite{Park:2018vci} is vital for computing the one-particle-irreducible (1PI) action. Our approach also sheds light on the subtle and difficult issue of gauge choice (in)dependence of the 1PI action \cite{Park:2018vci}.  
%%%%%%%%%%%%%%%%%%%%%%%%%%

In the foliation-based quantization (FBQ) \cite{Park:2014tia,Park:2014qoa,Park:2015qxa} it is the Lagrangian counterpart of the Hamiltonian constraint that leads to the physical state condition (PSC). More specifically, the lapse field equation of the Lagrangian Arnowitt-Deser-Misner (ADM) formalism was imposed as the physical state constraint in the previous sequels. In the present work the condition is alternatively derived from the gauge choice-independence of scattering amplitudes, after examining the gauge invariance issue of the boundary terms. The derivation of the PSC in this manner has two advantages over the previous. Firstly, the setup is manifestly 4D-covariant (other than splitting the gauge parameter $\e_\m$ into $\e_\m=(\e_{i},\e_3)$) because we do not, for the main task, resort to the ADM formalism. Secondly, the whole procedure is entirely within more established practice of quantum field theory: previously, the physical state condition was derived with hindsight of Dirac's method of quantizing a constrained system. Here, the condition is derived based on the conceptually more rudimentary requirement of gauge choice-independence of a scattering amplitude.

As an application of our quantization approach, we have recently tackled \cite{Park:2021ohu} the cosmological constant (CC) problem. Here we extend the one-loop analysis therein to two-loop. The CC problem was originally formulated in \cite{Weinberg:1988cp} with a generic system that contains a massive field whose contribution to the vacuum energy vastly exceeds the observed value of the CC. A good example is a loop contribution of the Standard Model (SM) Higgs field. The vacuum energy is defined as a minimum of the effective potential. In vacuum energy computation both the ultraviolet (UV) and infrared (IR) structures play roles. To some extent the UV and IR contributions to the vacuum energy are intertwined. Given that renormalization procedure is involved, the relevance of the UV structure is evident. The relevance of the infrared structure is subtler. As demonstrated in Casimir energy analysis (see, e.g., the account in \cite{Schwartz}), it is necessary to pay close attention to the infrared structure for proper evaluation of vacuum energy. (In the Casimir case this is often done by employing an infrared regulator of a finite-size box in momentum cutoff regularization.) We believe that the lesson learned from the Casimir case should be valid more generally: the vacuum energy of a system should be determined essentially by the low-energy sector of the theory, thus a meticulous description of the structure is desirable. Once temperature enters one deals with three different scales: the renormalized mass, the artificial energy scale introduced in dimensional regularization, and the temperature scale. As reviewed in \cite{Park:2021ohu}, convergence of perturbation theory dictates that these scales be on the same orders of magnitudes as one another. We show that there exists an optimal perturbation theory (OPT) \cite{Stevenson:1981vj} procedure that quantitatively enforces this qualitative requirement.

In another line of research we examine the issue of quantum electrodynamics (QED) asymptotic freedom \cite{Toms:2010vy} in a finite-temperature setup. Unlike our initial impression, it is likely that one must undertake the whole renormalization procedure, including other coupling constants, in order to properly investigate the potential asymptotic freedom at zero- or finite- temperature. This complication notwithstanding, QED asymptotic freedom remains a reasonable possibility.

\vspace{.3in}
The rest of the paper is organized as follows.
\vspace{.1in}

\ni In section 2 we examine the ordinary and large gauge symmetries, gauge-fixing, and residual gauge invariance. We note that the sector with a Dirichlet boundary condition should merely account for a `ground state' in the tower of the Hilbert space of the states arising from all possible boundary conditions. Non-Dirichlet boundary conditions are clearly motivated by large gauge transformations (LGTs). We discuss where the large gauge symmetry stands in the whole procedure of determining the PSC. We show that independence of the S-matrix under the residual symmetry leads to the physical state condition. Where useful, we use analogies with string theory. For instance, the ADM Lagrangian approach is analogous to lightcone string quantization, whereas the present covariant quantization is analogous, to some extent, to old covariant quantization. Section 3 is devoted to finite-temperature vacuum energy analysis. In section 3.1 we start by reviewing the zero-temperature CC problem. In section 3.2, we extend it to a finite-temperature setup. With the quantized metric contribution understood in \cite{Park:2021ohu}, we focus on flat space analysis of a real Higgs-type scalar system in a flat background. We recap two-loop effective potential computation in the refined background field method. In addition to the standard resummation, a non-perturbative technique of the so-called optimal perturbation theory was introduced in the literature for improved convergence. The way of implementing OPT is not unique. Our goal is to show that there exists a variant OPT procedure that allows one to avoid the CC problem. In section 4, we take an Einstein-scalar system in the finite-temperature framework of quantized gravity to examine the possibility of asymptotic freedom of QED proposed in \cite{Toms:2010vy}. In spite of the debates in the literature, we conclude that QED asymptotic freedom remains a reasonable possibility. In section 5 we conclude with a summary, implications of our results, and future directions. A glossary of some terms are given in Appendix A.

\section{Gauge symmetries, fixing, and PSC}

A crucial initial step in covariant gravity quantization involves handling of the constraints.
Since a gravity system is a gauge system, its quantization can be dealt with, in part, by Dirac's method (see, e.g., \cite{Weinberg} for a review), according to which second-class constraints can be formulated by Dirac brackets. The brackets are supplanted by the corresponding commutators at the quantum level. The core difficulty in quantization lies in first-class constraints. As explicitly demonstrated in recent works \cite{Park:2014tia,Park:2015ota,Park:2016zgt,Park:2018vci,Park:2019amz}, a first-class constraint can be taken care of by fixing the gauge symmetry that it generates \cite{Weinberg}. More specifically, the following was done \cite{Park:2014tia} in the {\em Lagrangian} ADM setup: with the lapse function and shift vector non-dynamical, their field equations were imposed as constraints upon gauge-fixing the lapse function and shift vector. This was in addition to bulk gauge-fixing (e.g., by the de Donder gauge). It was essentially the lapse field equation - the counter-part of the Hamiltonian constraint in the ADM {Hamiltonian} formalism (see \cite{Kiriushcheva:2010ycc} for a critical review of the Hamiltonian formalism of general relativity) - that led to reduction of the physical states. Although the ADM setup provides a convenient arena for and elucidates certain aspects of the quantization, one loses manifest 4D covariance. As we show in section 2.2, however, there is an alternative method of obtaining the physical state condition while maintaining the 4D covariance - which is almost always useful - in the intermediate steps: the condition is derived from gauge choice independence of scattering amplitudes, after examining the gauge-invariance issue of the boundary terms.\footnote{This is another case of the observation made in \cite{Park:2015xoa}: consideration of boundary physics leads to the same result obtained by considering bulk physics.}

We restrict our discussion to an asymptotically flat spacetime to prevent the analysis from getting too abstract, although the formalism may well be potent enough to cover more generic spacetimes. Furthermore, many interesting geometries are classified as an asymptotically flat spacetime. The advantage of considering an asymptotically flat spacetime is that it allows one to introduce a radial coordinate and, as we will discuss, one can do things more covariantly without resorting to the ADM formalism. We still split the gauge parameter $\e_\m$ into 
\be
\e_\m=(\e_{i},\e_\3)
\ee
with $x^{\m=3}\equiv r$, the radial direction, and ultimately focus on $\e_\3$.

In section 2.1 we start by reviewing the central idea behind the FBQ approach. We review aspects of the `small' (i.e., ordinary) and large gauge transformations. Although the asymptotic symmetry should in general be larger than the large gauge symmetry, we use the latter to be specific. The discussion will remain valid even if one takes the asymptotic symmetry. Some key ideas were explained in detail in the previous review \cite{Park:2019amz}; here the focus is on the latest developments. The `small' gauge transformation is redundancy of the degrees of freedom whereas the large represents part of the moduli (a collection of inequivalent vacua) of the theory \cite{Harvey:1996ur}\cite{DiVecchia:1998ky}. The reason for contrasting the {\em large} gauge symmetry with the ordinary gauge symmetry is to bring out precisely which symmetry is responsible for the PSC: it is part of - viz., residual symmetry of - the {\em ordinary} gauge symmetry whose handling in the manner described below leads to the PSC. With the preliminary discussions in section 2.1, we derive in section 2.2 the physical state condition \rf{psc} by requiring gauge-choice independence of a scattering amplitude under the residual symmetry.

\subsection{Review of gauge symmetry and its fixing}

The central idea on which the FBQ hinges is renormalizability of the physical sector associated with a 3D hypersurface in an asymptotic region. How does the reduction to the 3D hypersurface come about? As well known in gravitational (as well as non-gravitational gauge) theories, covariant gauges do not entirely exhaust the gauge freedom, but instead leave measure-zero (i.e., 3D) residual gauge redundancy. The key observation that led to the new approach to gravity quantization \cite{Park:2014tia} was that  suitable and complete gauge-fixing of 4D diffeomorphism and its residual symmetry leads to reduced support of the physical spectrum. More specifically, gauge-fixing of the 3D residual symmetry reduces the support of the physical states onto a holographic screen, a 3D hypersurface at an asymptotic location.

%%%%%%%%%%%%%%%%%%%%%%%%%%%%%%%%%%%
Although 4D diffeomorphism is something well-established, there are several fine but nonetheless crucial issues that one must carefully discern. This is partially due to the fact that the two types of gauge transformations, the ordinary (or `small') and the large, are `tangled'. It takes some care to disentangle the two, a required step to determine the physical states in the present approach. Also, the observer-dependent effects are tied with the gauge transformations and resulting foliations \cite{Freidel:2016bxd}\cite{James:2016zha}. Let us note that there are two kinds of residual symmetry, both of which correspond to each type of the gauge symmetry: the first is the residual symmetry of ordinary diffeomorphism.
The residual symmetry of the second kind is associated with the large gauge symmetry. (For convenience we are viewing, at the moment, the large symmetry as the residual symmetry of the small and large symmetries combined.) It will be lucrative to invoke analogies with string theory: the large gauge symmetry is an analogue of modular group whereas the residual symmetry is analogous to conformal Killing group.

As for the ordinary gauge symmetry, it will be useful to briefly remind us of the derivation of the PSC in the Lagrangian ADM setup before getting to the quantitative details of the 4D-covariant derivation. The residual symmetry associated with $\e_{i}$ is used to gauge-fix the shift vector, as analyzed in the earlier sequels. With this, one can focus on the gauge parameter of the form 
\be
\e^\m=(0,0,0,\e^{\3}(t,r,\th,\f))
\ee
with a property
\be
\e^\3\equiv \e^{\m=3}(t,r,\th,\f)\ra 0\;\; \mbox{as}\;\, r\ra \infty. 
\ee
The residual symmetry of ordinary gauge symmetry that leads to the physical state condition is one associated with $\e_\3$. One first fixes the lapse by using the residual 3D symmetry generated by $\e_\3$.\footnote{For a simpler background, such as a Schwarzschild background, the lapse function need not be fixed - it is determined while solving the shift vector constraint. In general, one should use the symmetry to fix the lapse.} 
The lapse equation of motion is a first-class constraint and generates, in the Dirac formulation, a translation along the `time-', i.e., $r$- direction. Thus an $r$-translation is part of the gauge redundancy (but not part of the moduli). The lapse equation of motion as a constraint reduces the support of the physical states to a hypersurface at the asymptotic region of $r=\infty$. 
%%%%%%%%%%%%%%%%%%%%%%%

It is also useful to distinguish the above residual symmetry from the conformal-type symmetry contained in the diffeomorphism. The latter takes a special form and is associated with the trace part of the fluctuation metric. This can be seen by recasting the diffeomorphism transformation with a parameter $\xi^\m$,
$\d g_{\m\n}=\nabla_\m \xi_\n+\nabla_\n \xi_\m$,
into the form
\bea
\d g_{\m\n}=\fr12 (\nabla_\k \xi^\k)g_{\m\n}+(Lg)_{\m\n} \la{diffeo}
\eea
where $(Lg)_{\m\n}$ denotes the traceless part of the Lie derivative
\bea
(Lg)_{\m\n}\equiv \nabla_\m \xi_\n+\nabla_\n \xi_\m-\fr12 (\nabla_\k \xi^\k)g_{\m\n}.
\eea
The first term in \rf{diffeo} takes the form of a conformal transformation. This symmetry must be removed by gauge-fixing of the trace piece of the fluctuation metric \cite{Park:2015xoa}.

Remarks on the large gauge symmetry are in order. As previously mentioned, a Dirichlet boundary condition should cover merely the measure-zero subset of possible boundary terms and conditions. Recall that the Dirichlet boundary condition has a special status in that it is imposed when defining the canonical momenta. (Once the momenta are defined, one may consider other types of boundary conditions.) Since an LGT will not, generally speaking, preserve the boundary conditions (because, for one thing, it will not preserve the momenta), different boundary conditions should be viewed as different sectors of the theory. For this reason the large gauge symmetry is analogous to global symmetry or moduli. For us an LGT will be an asymptotically non-vanishing 3D transformation in the $(t,\th,\f)$ space. (It is also the degrees of freedom associated with the reduced action: the large gauge symmetry must be non-perturbative degrees of freedom of the reduced action obtained in \cite{Park:2018xtt}.) Now consider the 4D action with boundary terms. An LGT mixes the time and spatial coordinates and will not, in general, leave the content of the original boundary condition invariant. In addition, due to the mixing of the coordinates, observer-dependent effects will enter \cite{Freidel:2016bxd}\cite{James:2016zha}.

Another not unrelated key ingredient in deriving the PSC is careful treatment of the boundary dynamics, including boundary terms with the corresponding boundary conditions. The boundary terms are important not only on their own, but also for identifying the physical states. Although Dirichlet boundary conditions are widely used in gravity (as well as other field theories), it has been shown that the Hilbert space must be extended so as to include non-Dirichlet sectors. The point is that exclusive imposition of a Dirichlet boundary condition cannot be justified since an ordinary gauge transformation does not preserve them (more precisely, the content of the boundary condition, though the form of the boundary term should be invariant). What is missed by restricting to the Dirichlet boundary condition is the entire boundary dynamics. The sector with the Dirichlet boundary condition should only account for a `ground state' of the tower of the Hilbert space of states coming from all possible boundary conditions. For this reason, the physical state condition should be derived in the setup of an extended Hilbert space.

\subsection{Alternative derivation of PSC}

With the preliminary in section 2.1, we are ready to derive the PSC from gauge choice-independence of a scattering amplitude. As pointed out before, the Lagrangian ADM method is analogous to string theory lightcone quantization in that one maximally exploits gauge-fixing. In old covariant quantization, on the other hand, the physical states are realized through imposition of appropriate constraints. Here, we do something similar in spirit: we impose the `lapse constraint' without explicitly fixing the `lapse'. (Quotation marks since the analysis is conducted in the 4D-covariant frame work.) This may be regarded as a `cohomological' way of obtaining the physical states. The physical states must be invariant under $r$-translation, which is part of the residual symmetry. In general, a bulk state cannot satisfy this condition, and one must turn to a state that has support on an asymptotic boundary. The weaker form of the physical state condition \rf{psc} can be derived by carefully examining the Dirichlet and Neumann boundary conditions, as we now do.

Let us start with an Einstein-Hilbert (EH) action with a York-Gibbons-Hawking (YGH) boundary term:
\be
S_{EH+YGH}=S_{EH}+S_{YGH}
\ee
\be
S_{EH}\equiv \int d^4x \sqrt{-g}\; R\quad,\quad S_{YGH}\equiv 2\int_{\pa {\cal V}} d^3x \sqrt{|\g|}\, \ve K.
\la{ghyt}
\ee
where $\g_{\m\n}$ denotes the induced metric on the boundary ${\pa {\cal V}}$; $\ve$ takes $\ve=-1$ for the usual foliation with the genuine time coordinate $x^{\m=0}=t$, whereas it takes $\ve=1$ for the $r$-foliation.
Let us first quickly remind us of the standard procedure of obtaining the equation of motion with the Dirichlet boundary condition. Variation of the Einstein-Hilbert action consists of bulk terms and a boundary term. The former leads to the equation of motion; the latter comes from
\bea
\int_{{\cal V}} \sqrt{g}\; g^{\m\n}\d R_{\m\n}&=&\int_{{\cal V}} \sqrt{g}\;\nabla^\m
\Big[\nabla^\n \d g_{\m\n}-g^{\r\s}\nabla_\m \d g_{\r\s} \Big]  \nn\\
&=&\int_{\pa{\cal V}} \sqrt{g}\;n^\m
\Big[\nabla^\n \d g_{\m\n}-g^{\r\s}\nabla_\m \d g_{\r\s} \Big]  \nn\\
&=&\int_{\pa{\cal V}} \sqrt{g}\;n^\m g^{\n\k}
\Big[\nabla_\k \d g_{\m\n}-\nabla_\m \d g_{\n\k} \Big]
\la{bdterms}
\eea 
where $\d$ denotes an arbitrary variation (as opposed to the symmetry variation, $\d_\e$, below). By employing the standard splitting $g^{\n\k}=\ve n^\n n^\k+\g^{\n\k}$, $n_\m$ being the unit normal to ${\pa {\cal V}}$, and noting the (anti)symmetry in $(\m,k)$, one gets
\bea
&=&\int_{\pa{\cal V}} \sqrt{g}\;n^\m \g^{\n\k}
\Big[\nabla_\k \d g_{\m\n}-\nabla_\m \d g_{\n\k} \Big]. \la{bdterms2}
\eea 
The second term inside the parentheses is canceled against $\d S_{GHY}$; requiring vanishing of the first term  is the Dirichlet boundary condition.

As for the `cohomological' determination of the {\em physical states}, one must consider a symmetry variation instead, and establish invariance of the action under the residual symmetry of the ordinary gauge symmetry. Then the physical states will be ones invariant under $r$-translation. It would be ideal to consider the most general boundary terms if such terms were known. This not being the case, we are content to demonstrate invariance\footnote{This is not without a subtlety; see the comments at the end.} for the Dirichlet and Neumann sectors. For the Dirichlet case, the exact same steps above apply when considering a symmetry variation instead of an arbitrary variation. One just needs to use the gauge parameter $\e^\m$ such that $\e^\m \ra 0$ as $x^{\mu=3}\equiv r\ra \infty$, in order for the metric variation $\d_\e g_{\m\n}$ to preserve the Dirichlet boundary condition. As for the Neumann case we consider the action of $S_{EH}$ alone without $S_{YGH}$ \cite{Krishnan:2016tqj}. We show that the action with the Neumann boundary condition is invariant so far as one imposes the traceless condition. The second term inside the square parenthesis in \rf{bdterms} vanishes due to the traceless condition: 
\bea
\int_{\pa{\cal V}} \sqrt{g}\;n^\m g^{\n\k}
(-)\nabla_\m \d g_{\n\k}=0.
\eea 
As for the first term, note that
\bea
	\nabla^\n \d_\e g_{\m\n}=g^{\n\n'}\nabla_{\n'} \d_\e g_{\m\n}=g^{\n\n'} \nabla_{\n'} \mathscr{L}_\e g_{\m\n}
	=g^{\n\n'} \Big([\nabla_{\n'}, \mathscr{L}_\e] g_{\m\n}+  \mathscr{L}_\e \nabla_{\n'} g_{\m\n}\Big)
\la{clc}
\eea
where $\mathscr{L}$ denotes a Lie derivative. The second term in the far right-hand side of \rf{clc} vanishes; the first term vanishes as well since $[\nabla_{\n'}, \mathscr{L}_\e]= \fr{\pa \e^\k}{\pa x^{\n'}}\nabla_\k$ \cite{Kobayashi}\cite{Park:2014qoa,Park:2015qxa} acting on the metric vanishes. 

Let us pause and recapitulate. With the YGH boundary term added, one gets invariance, $\d_\e S_{EH+YGH}=0$, once one imposes the Dirichlet boundary condition $\g^{\n\k}\nabla_\k \d_\e g_{\m\n}=0$. It has also been shown that with the Neumann boundary condition, namely, without the YGH boundary term, one gets $\d_\e S_{EH}=0$ as far as one imposes the traceless condition. The point is that if one does not remove the traceless mode, one can nevertheless achieve the invariance of the action including the boundary terms by imposing the Dirichlet boundary condition. What we have just shown above is that once one imposes the traceless condition, the gauge variation of $S_{EH}$ vanishes without the use or presence of the YGH term.\footnote{One can alternatively proceed with the boundary expression in terms of $K$, the trace of the second fundamental form. The invariance just established implies $\d_\e K=0$. This means that the Dirichlet or Neumann is preserved by a gauge transformation.}

Let us now employ the 3+1 splitting of $\e^\m$ in the 4D approach and focus on $\e^\3$. Denote by $Q$ the transformation generated by the `time' translation: $\d_\e$ is promoted, at the quantum level, to the corresponding charge operator, $Q_\e$. Let us briefly pause and translate things into the ADM Lagrangian perspective. $Q_{\e^\3}$ is nothing but the lapse function constraint (with gauge parameter $\e^\3$ included). The connection between $Q_{\e^\3}$ and the lapse field equation is that the lapse function may be gauge-fixed by the symmetry generated by $Q_{\e^\3}$. This means that in the Lagrangian ADM formalism $Q_{\e^\3}$ is the lapse field equation.

Finally, the invariance that we have established amounts, at the quantum level, to
\be
	[Q_{\e^\3},S_{EH+YGH}]=0
\ee	
for the Dirichlet case, and
\be
	[Q_{\e^\3},S_{EH}]=0
\ee	
for the Neumann case. Therefore the physical content of the theory is determined by the physical state condition in analogy with a cohomological case:
\be
	Q_{\e^\3}|\mbox{physical state}>=0. \la{psc}
\ee

\vspace{.in}
To end this section, we comment on the aforementioned subtlety in establishing the invariance of the action with the boundary terms. The subtlety is present in any boundary condition; we illustrate it with the Dirichlet boundary condition by taking the first term of \rf{bdterms2}. Although the first term does not vanish in the transformed coordinate system with the Dirichlet boundary condition {\em that is natural in the new coordinates}, one sets this aside and achieves the invariance up to this point. The `deficit' is subsequently addressed through the channel of the observer-dependent effects \cite{Freidel:2016bxd}\cite{James:2016zha,Park:2018xtt}.

%%%%%%%%%%%%%%%%%%%%%%%%%%%%%%%%%%%%%%%
%%%%%%%%%%%%%%%%%%%%%%%%%%%%%%%%%%%%%%%
\section{Vacuum energy in finite temperature}
%%%%%%%%%%%%%%%%%%%%%%%%%%%%%%%%%%%%%%%
%%%%%%%%%%%%%%%%%%%%%%%%%%%%%%%%%%%%%%%

When the characteristic scale of the theory under consideration, say, the electroweak scale, is much higher than the `room' temperature or the temperature of cosmic microwave background (CMB), it is standard practice to employ zero-temperature field theory. Although employing zero-temperature field theory may seem innocuous, our analysis indicates otherwise: finite-temperature effects reveal, when properly taken into account, how to carry out perturbation theory in a `natural' manner.\footnote{Interestingly, the role of thermodynamics in determining vacuum energy has recently been explored in \cite{Ryskin:2014pva,Ryskin:2020yod}, the works that I became aware after completion of \cite{Park:2021ohu}.} In this section we extend the analysis in \cite{Park:2021ohu} to two-loop and show that finite-temperature effects are the key to avoiding the CC fine-tuning problem. We analyze the CC problem by taking an Einstein-scalar system with a Higgs-type potential. Dimensional regularization, which has a well-known advantage in dealing with a gauge system, is employed.

With temperature present one deals with three different scales: the renormalized mass, the artificial energy scale introduced by dimensional regularization, and the temperature scale itself.  A potential danger in disregarding (in spite of a low temperature) the finite-temperature effects and turning to zero-temperature theory is hinted at by well-known patterns in perturbation series:  logarithms of ratios of these scales appear in the series. As lucidly reviewed in \cite{Park:2021ohu}, convergence of perturbation theory makes it necessary for these scales to be on the same order of magnitude as one another; it is not at all clear whether taking zero-temperature field theory with a large renormalized scalar mass would yield results that are consistent with ones obtained by taking a low temperature limit of the system at finite temperature. What we convey below is that although the two approaches should be compatible, there is a (bigger) price to pay for the zero-temperature approach: a posteriori, the unnaturalness of the approach manifests as the CC problem.

Regardless of justification for applying zero-temperature field theory to a low but nonzero temperature system, it would be valuable to have techniques that can cover physics from near Planck temperature to the CMB temperature. An obstruction to such a full-range description is finite-temperature infrared problems. The most serious among those is the `Linde problem' \cite{Linde:1980ts} at QCD-scale temperature. Resummation and various non-perturbative techniques were introduced to deal with the problem (see, e.g., \cite{Karsch:1997gj} and \cite{Andersen:2001ez}). The focus of the present work is low temperature, the temperature of CMB. In particular, we explore reformulation of the CC problem - which was originally formulated in the zero-temperature setup - as a zero-temperature limit of a {\em finite}-temperature setup. In the main body we show that the finite-temperature effects are in fact crucial - they allow one to avoid the CC fine-tuning problem once the convergence property of the perturbation series is improved through a variant of optimal perturbation theory (OPT).

The analysis in the present work has the following components: UV divergence removal in finite temperature, OPT-improved resummation and renormalization, and the house keeping setup of quantized gravity. Ultraviolet renormalization at finite temperature is guaranteed if the zero-temperature renormalizability is established, and plays an important role - similar to that in Casimir energy computation - through renormalization conditions. In the finite-temperature literature, resummation was introduced long ago to mitigate the temperature-induced divergences in the infrared regime. The convergence properties can be further improved with a touch of non-perturbative techniques, OPT. The OPT that we implement in this work is a relatively minor, but nonetheless crucial, variation of the widely used kind. It is these OPT-organized finite-temperature effects that ultimately turn out to be central to the proposed resolution of the CC problem.

We show in the main body that the optimized renormalized mass turns out to be essentially the temperature. We believe that this allows one to identify (and solve) the cosmological constant problem at its root. With the renormalized mass determined, the following task still remains: the zero-temperature theory has been quite successful for other purposes, and there, the renormalized mass is taken quite close to the pole mass value. In the SM case, the renormalized Higgs mass is taken to be close to the physical value, 125 GeV, within a few percent. If one now wants to take the renormalized mass to be around the CMB temperature, which is much smaller than the pole mass, one must maintain that the perturbation theory with the corresponding mass preserves the success of the SM. As analyzed in \cite{Park:2021ohu} renormalization invariance of physical quantities - which implies a certain additional resummation identified therein - can be invoked to confirm that.

In section 3.1 we review the refined BFM computation of the effective potential and CC problem. We do this for  zero-temperature as in the original work of \cite{Weinberg:1988cp}. In section 3.2 we formulate the problem in the finite-temperature context. Our analysis borrows \cite{Park:2021ohu} a high-temperature expansion of the effective potential, in spite of the fact that the temperature considered is that of CMB. 
In the effective potential analysis one encounters a novelty: the potential becomes complex, indicating instability of the vacuum \cite{Park:2021ohu}.

%%%%%%%%%%%%%%%%%%%%%%%%%%%%%%
%%%%%%%%%%%%%%%%%%%%%%%%%%%%%%
\subsection{Review of CC problem}
%%%%%%%%%%%%%%%%%%%%%%%%%%%%%%
%%%%%%%%%%%%%%%%%%%%%%%%%%%%%%

In this subsection we present a brief and streamlined review of zero-temperature computation of the effective potential in the refined BFM \cite{Park:2014noa}\cite{Park:2016zgt}\cite{Park:2018vci,Park:2019amz}, thereby setting the stage for the two-loop analysis in section 3.2. We give an account of the CC problem in dimensional regularization, which is also employed in the two-loop analysis. (Recall that momentum cutoff regularization was employed in the original observation of the CC problem in \cite{Weinberg:1988cp}.) Several different non-perturbative techniques were put forth to deal with the infrared divergence problem; we adopt a version of OPT in which the renormalized mass itself plays the role of a variational parameter.

The CC problem was originally established by considering quantized matter fields in a flat spacetime at {\em zero} temperature. It will be useful to review the CC problem in the same setup (but with dimensional regularization), prior to finite-temperature treatment. Ultimately, the entire two-loop analysis will have to be founded on a setup of quantum gravity: renormalization of CC would not have a rationale were it not for quantized gravity. Happily, the analysis can be carried over to the quantized gravity setup - where the CC comes to have its proper meaning - without any major difficulty \cite{Park:2021ohu}.

The refined - as opposed to the conventional - BFM is employed for two reasons: firstly, it makes it clear that, for the vacuum energy, what one needs is an onshell value of the {\em offshell} potential. For this one should solve the quantum-corrected offshell potential for its minimum. Employing the refined BFM avoids, as we elaborate below, possible confusion on the onshell vs. offshell issue. Secondly, whereas employing the refined BFM is a matter of convenience for the matter sector, this is not the case for the graviton sector. For the graviton sector, it is necessary to employ the refined BFM to ensure covariance of the effective action \cite{Park:2014noa}\cite{Park:2015ota}.

Consider the following renormalized scalar action in a flat background at zero temperature:
\bea
S(\z) 
&=& -\int d^4x\;\left[ \fr12 \pa_\m \z \pa^\m \z +\fr14 \tilde{\l} \Big(\z^2 +\fr{\n^2}{\tilde{\l}}\Big)^2\right]
\la{flatsact}
\eea
where we have defined
\be
\tilde{\l}\equiv \fr16 \l.
\ee
Note we have adopted the complete-square form of the potential, instead of the more usual $V=\fr12 \n^2 \z^2+\fr14 \tilde{\l} \z^4$. As addressed in more detail in \cite{Park:2021ohu}, whether one should use the complete-square form or the more usual form is not part of the CC problem; it is an independent problem whose answer must ultimately be given by experiment. Our goal of establishing the absence of the fine-tuning-problem can be more handily achieved with the complete-square form. With it, the classical potential vanishes onshell - namely, once one sets $\z$ to
\be
\z_0^2=-\fr{\n^2}{\tilde{\l}}. \la{czvev}
\ee  
Compare the conventional and refined BFMs. In the former one shifts the field according to
\be
\z\ra \z+\z_0;  \la{bfmcs}
\ee 
which yields
\bea
&&\hspace{.6in}S(\z+\z_0)= -\int d^4x\;\fr14 \tilde{\l} \Big(\z_0^2 +\fr{\n^2}{\tilde{\l}}\Big)^2 \nn\\
&&\hspace{-.3in}-\int d^4x\; \Big[\fr12 \pa_\m \z \pa^\m \z +\fr12 \n^2 \z^2\Big]-\int d^4x\; \tilde{\l}\Big(\fr32 \z_0^2 \z^2+ \z_0\z^3+\fr14  \z^4\Big) \la{esa} 
\eea 
where the $\z$-linear term has been omitted as usual. The effective potential is obtained by integrating out $\z$ running on the loops with $\z_0$s sitting as external legs in a 1PI diagram.  In the refined BFM, on the other hand, the shift is taken to be
\be
\z\ra \z+\tilde{\z},\quad  \tilde{\z}\equiv \z_c+\xi. \la{srbfm}
\ee
$\z_c$ denotes a classical solution of the original action; $\xi$ is the background field.\footnote{This is the case in the so-called second-layer perturbation. In the first-layer perturbation $\tilde{\z}$ is taken as the background field \cite{Park:2018vci}.} In the conventional BFM, $\z_0$ is taken to be a constant (as demonstrated in \rf{bfmcs}) when one is interested in the effective potential and its vacuum solution. In the refined BFM, a solution $\z_c$ will of course be a function of the coordinates in general. With the shift in \rf{srbfm} one gets
\bea
&& S(\z+\tilde{\z})= -\int d^4x\;\Big[\fr12 \pa_\m \tilde{\z} \pa^\m \tilde{\z}+\fr14 \tilde{\l} \Big(\tilde{\z}^2 +\fr{\n^2}{\tilde{\l} }\Big)^2  \Big]\nn\\
&&\hspace{-.27in} -\int d^4x\; \Big[\fr12 \pa_\m \z \pa^\m \z +\fr12 \n^2 \z^2\Big]-\int d^4x\; \tilde{\l}\Big(\fr32 \tilde{\z}^2 \z^2+ \tilde{\z}\z^3+\fr14  \z^4\Big). \la{esaappq} 
\eea
If one is interested in the effective potential as opposed to the effective action, the conventional BFM becomes equivalent to the refined upon identifying $\z_0=\tilde{\z}$; which BFM to employ is a matter of convenience in this sense. However, things are much subtler in the gravity sector: it is only the refined BFM that yields the correct results. In dimensional regularization, one introduces a scale parameter $\m$:
\bea
&& S(\z+\tilde{\z})= -\int d^4x\;\Big[\fr12 \pa_\m \tilde{\z} \pa^\m \tilde{\z}+\fr14 \tilde{\l} \m^{2\ve}\Big(\tilde{\z}^2 +\fr{\n^2}{\tilde{\l} \m^{2\ve}}\Big)^2  \Big]\nn\\
&&\hspace{-.3in} -\int d^4x\; \Big[\fr12 \pa_\m \z \pa^\m \z +\fr12 \n^2 \z^2\Big]-\int d^4x\; \tilde{\l}\m^{2\ve}\Big(\fr32 \tilde{\z}^2 \z^2+ \tilde{\z}\z^3+\fr14  \z^4\Big). \la{esaappq2} 
\eea
The effective action can be computed by organizing the diagrams in order of increasing number of external $\tilde{\z}$-fields. Since we are interested in the potential part of the effective action, for which one can treat $\tilde{\z}$ as constant, it is more efficacious to collect the terms quadratic in $\z$ and treat them as part of the kinetic term:
\bea
\exp \Big(i\G^{\mbox{1-loop}}(\tilde{\z}) \Big)=\int d\z\; \exp\Big[-i \int d^4x\; \Big(\fr12 \pa_\m \z \pa^\m \z +\fr12 M^2(\n,\tilde{\z}) \z^2\Big)\Big] 
\eea 
where 
\be
M^2(\n,\tilde{\z}) \equiv \n^2+3\tilde{\l} \tilde{\z}^2. \la{Mdef}
\ee
Note that for the one-loop effective potential, only the kinetic terms contribute, as indicated above. One gets 
\be
V^{\mbox{1-loop}}=-\fr{i }{2(2\pi)^4}\int d^4p\;\ln\fr{i}{\pi}\Big[{p^2+M^2(\n,\tilde{\z})}\Big].
\la{epi}
\ee
Combining the tree and one-loop results yields
\bea
V(\tilde{\z}) &=& \fr14 \tilde{\l} \Big(\tilde{\z}^2 +\fr{\n^2}{\tilde{\l}}\Big)^2 - \fr{i}{2(2\pi)^4}\int d^4p\;\ln\Big[\fr{i}{\pi}({p^2+M^2(\n,\tilde{\z})})\Big].
\eea 
In dimensional regularization:
\bea
V^{\mbox{1-loop}} 
&=& -\fr{1}{32\pi^2}\Big(\fr{1}{2\ve}+\cdots\Big) \,M^4(\n,\tilde{\z}). \la{1lpp}
\eea
The $\fr1{\ve}$ term must be subtracted out by a CC counter-term. To see the CC problem, it is sufficient to consider the leading order: the minimum of the one-loop corrected potential occurs at 
\be
\tilde{\z}_m= \sqrt{-\fr{\n^2}{\tilde{\l}}}+{\cal O}(\hbar). \la{vevexp}
\ee
The CC, i.e., the value of the potential evaluated at $\tilde{\z}_m$ above, is on the order of
\be
\sim M^4\Big|_{\tilde{\z}= \tilde{\z}_m}\sim \n^4. \la{Mtf}
\ee
In Standard Model, the renormalized mass in the modified minimal subtraction ($\overline{\mbox{MS}}$) scheme is determined at the end by requiring the pole of the Green’s function to take the physical value, or 125 GeV in the case of the Higgs field. The renormalized mass turns out to be quite close to the pole masses, usually within a few percent. Upon substituting the physical value of $\n$, i.e., the value of $\n$ corresponding to the physical value of the Higgs mass, the result above leads to a CC value enormously bigger than that of the observed. This is the CC problem: a highly fine-tuned renormalization procedure is required to bring the theoretical one-loop value down to the much smaller observed value.

\vspace{.1in}

As reviewed above, the CC fine-tuning problem is quite generically present as long as the renormalized mass (its fourth power, to be precise) is much larger than the observed value of the CC. One is then naturally led to the question of whether or not there exists a rationale by which one can employ a renormalized mass of the Higgs field that is far smaller than 125 GeV and carry out the renormalization program. We point out two (relatively) well-known facts as a positive indicator toward such a program. One is flexibility in renormalization schemes, also known as renormalization conditions or subtraction schemes. After UV regularization one must subtract out the infinite part and fix the finite part of the vacuum energy. In $\overline{\mbox{MS}}$ scheme one removes essentially only $\fr{1}{\ve}$ part. This fixes the finite part; at this point the renormalized mass is yet to be determined. It is determined by matching the pole value of the 2-point function with the physical value of the field, 125 GeV for the Higgs field. In the proposed new scheme it is the value of the renormalized mass, instead of the finite part, that is first fixed (to be on the order of the temperature). Subsequent matching with the physical mass then determines the finite part. The other sign is one associated with the presence of temperature. Once temperature enters, the zero-temperature setup becomes unsuited (which seems to manifest as the CC fine-tuning eventually). An indication of this comes from energy scalings in finite-temperature loop analysis. Recall that in zero temperature a loop analysis typically yields logarithmic factors such as $\ln\fr{m}{\m}$, where $m$ is the renormalized mass of the field and $\m$ the renormalization scale. For the benefit of convergence, it is necessary to choose $\m\sim m$. By the same token it will be necessary to take $\m\sim m \sim T$ once the temperature enters. In the present work this scaling is quantitatively achieved in the course of improving the perturbative analysis by optimal perturbation theory (OPT) after standard thermal resummation - we show that there exists an OPT procedure that enforces the scaling.

%%%%%%%%%%%%%%%%%%%%%%%%%%%%%%
%%%%%%%%%%%%%%%%%%%%%%%%%%%%%%
\subsection{finite-T analysis and resolution of CC problem}
%%%%%%%%%%%%%%%%%%%%%%%%%%%%%%
%%%%%%%%%%%%%%%%%%%%%%%%%%%%%%

For CC renormalization all of the fields, including the metric, must be quantized and their contributions to the CC counted. For the pure graviton sector, the structure of $n$-loop contributions with $n\geq 2$ have been analyzed in \cite{Park:2021ohu}. In this section we focus on the matter sector and conduct two-loop analysis of thermal effects by taking a gravity-Higgs system in a flat background. Consideration of a flat non-dynamical metric background is for simplicity, for one thing. Naively, the contributions of massive matter fields to the CC are expected to be larger than those of the gravitons. The variant OPT unravels, however, that the ultimate determining factor of the CC in the present scheme is the temperature.

We consider a slightly modified version of the widely-adopted OPT. In the widely-used version (see, e.g., \cite{Chiku:1998kd}), an artificial mass term is added and subsequently subtracted out. This is one way of ensuring that the artificial mass term would not have any effect on the full closed-form results. Although the mass term would not affect the full closed results, it, serving as a variational parameter, does improve finite-order analysis: in the present implementation of OPT, the renormalized mass itself will serve as the OPT parameter to be determined by principle of minimal sensitivity (PMS) \cite{Stevenson:1981vj}. As known in the context of the variational principle in quantum mechanics, there is no unique scheme for implementing the principle. For instance, the more variational parameters one introduces, the more accurate the approximation generally becomes. Our OPT is one that has an advantage of achieving avoidance of the CC problem. What is important for the CC problem is that such OPT exists.

Below, we first discuss several issues including justification and benefits of considering a flat spacetime. We then carry out two-loop analysis in a flat spacetime. We review computation of the potential with the standard resummation by carefully keeping track of the relevant structures. Our OPT is then implemented, and optimized renormalized mass is determined. One encounters the novelty mentioned in the introduction to this section in that the potential becomes complex.\footnote{Strictly speaking, the potential itself remains real even at two-loop. However, the vev of the scalar field becomes complex. It is expected that the complexity of the potential will become manifest at three-loop.}

\vspace{.2in}

Taking both finite-temperature (see, e.g., \cite{Kapusta}\cite{Bellac}\cite{Laine:2016hma} for reviews) and metric-loop effects into account requires intensive effort. To be entirely realistic, one would also have to consider an FRLW-type time-dependent background. Needless to say, doing all these at the same time should be a daunting task\footnote{In addition, there is a potential complication caused by the fact that our Universe was not always in equilibrium. It would be ideal if one could apply non-equilibrium thermodynamic QFT to the problem. However, it is not clear whether non-equilibrium thermodynamic QFT has been sufficiently developed for such a purpose. As a workaround, one may perform the present analysis for an epoch that was either in equilibrium (since our Universe has been mostly in equilibrium) or will be close to it (e.g., a future time when the Universe gets close to equilibrium heat death).}: one would first need to realize the FRLW background as a solution of the Einstein-scalar system. Although this may be possible in a certain series approximation, a closed analytic form of the solution would be desirable for ensuing analysis. Furthermore, the propagators in such a background would be highly involved even without the presence of temperature. One would be hampered by these technical complexities early on in the undertaking. 

Fortunately, however, the crux of the CC resolution can be captured by considering a scalar system in a flat spacetime \cite{Park:2016zgt}\cite{Park:2018vci}. A flat-spacetime analysis is in fact more than a toy model. This is especially so for UV divergences, since they come from  high-energy virtual particles. In other words, since the UV divergences originate locally from a short-distance, they are insensitive to the global geometry. Similarly, a finite-temperature theory can employ the same UV regularization as the zero-temperature theory. As for the things that depend on the infrared structure, the prime example of which is vacuum energy, one must in principle consider the actual background. As we show below the energy scale is correlated with the temperature. Ideally, one should thus use the actual curved background when the temperature becomes low. The difference between using the actual curved background and a flat one instead lies in finite parts. The finite parts can (and must) be adjusted by the renormalization conditions anyway.

We now come to the heart of section 3. Because we will be interested in temperature much lower than the electroweak (EW) scale, an obvious question is whether or not there would be any room for finite-temperature effects. The answer is affirmative, as we show. We conduct the standard resummation followed by variant OPT implementation. The one-loop observation in \cite{Park:2021ohu} that the CC problem is avoided is extended to two-loop.

\vspace{.2in}
The finite-temperature propagator associated with the action \rf{flatsact} is
\bea
<\z(x_1)\z(x_2)>=T\sum_n e^{-i\w_n(\t_1-\t_2)}\int \fr{d^3k}{(2\pi)^3}\fr{e^{i{\bf k}\cdot ({\bf x_1-x_2})}}{i (\w_n^2+{\bf k}^2+M^2)}  \la{tprop}
\eea
where $\w_n\equiv 2\pi Tn$ ($n=0,1,2...$) and $M^2\equiv \n^2$ before resummation. (See eq. \rf{Massterm} for comparison.) At one-loop, renormalization of mass and coupling is necessary. Renormalization of the constant part of the potential, $\fr{\n^4}{4\tilde{\l}}$, is also needed. By introducing the renormalization constants, $Z_0,Z_1,Z_2$, the bare action may be written as
\bea
&&S_B(\z)= -\int d^4x\; \Big[\fr12 \pa_\m \z \pa^\m \z + \fr12 Z_1\n^2 \z^2\Big]-\int d^4x\; \fr{Z_2\tilde{\l}}4  \z^4 -\int d^4x\;  \fr{Z_0\n^4}{4\tilde{\l}}.  \nn\\ \la{bare} 
\eea
Precise forms of the two-loop parts of the renormalization constant $Z$'s are not needed for the goal. By the same token it is not necessary to keep tract of the wave-function renormalization constant $Z_3$: since we employ $\overline{\mbox{MS}}$ scheme (initially), $Z_3$ is determined by offsetting the divergences remaining after being partially canceled by the other $Z$-constants' contributions. 

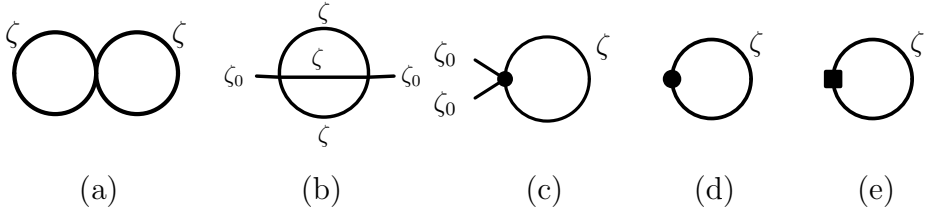
\begin{figure}[t]
	%	\begin{center}
	%%%%%%%%%%%%%%%%%%%%%%%%%%%%
	\hspace{.4in}
	\begin{fmffile}{earsm}
		\Scale[1.6]{
			\begin{fmfgraph*}(40,29)
				\\
				\fmfleft{i}
				\fmfright{o}
				\fmf{phantom,tension=11}{i,i1}
				\fmf{phantom,tension=11}{o,o1}
				\fmf{plain,left=1.0,tension=0.2}{i1,v1,i1}
				\fmf{plain,right=1.0,tension=0.2}{o1,v1,o1}
				%\fmf{dashes}{v1,v2}
				\fmf{phantom,label=\scalebox{.6}{${ \z}$}}{o,o1}
				\fmf{phantom,label=\scalebox{.6}{${ \z}$}}{i1,i}
			\end{fmfgraph*}
		}
	\end{fmffile}\;\;\;\;\;
	%%%%%%%%%%%%%%%%%%%%%%%%%%%%%%   
	\hspace{-.1in}
	\begin{fmffile}{sunsetm}
		\Scale[1.3]{
			\begin{fmfgraph*}(40,34)
				\fmfleft{i}
				\fmfright{o}
				\fmfforce{(.3w,0.3h)}{v}
				\fmf{plain,tension=20}{i,v1}
				\fmf{plain,tension=20}{v2,o}
				%	\fmf{plain,left,tension=0.4}{v1,v2,v1}
				\fmf{plain}{v1,v2}
				%	\fmfdot{v1,v2}		
				\fmflabel{\!\!\scalebox{.6}{${ \z_0}$}}{o} 	
				\fmflabel{\scalebox{.6}{${ \z_0\!\!}$}}{i} 		
				%	\fmf{plain,label=\scalebox{.7}{${ \z}$}}{v2,v1}
				%	\fmf{plain,label=\scalebox{.5}{${ \z}$}}{v1,v2,v1}
				\fmf{plain,left=1.1}{v1,v2}
				\fmf{plain,left=.9}{v2,v1}
				\fmf{phantom,left=0.73,label=\scalebox{.6}{${ \z}$}}{v1,v2}
				\fmf{phantom,left=0.6,label=\scalebox{.6}{${ \z}$}}{v2,v1}
				\fmf{phantom,left=0.6,label=\scalebox{.6}{${\z}$}}{v1,v}
				\fmf{phantom,left=0.6}{v2,v}
			\end{fmfgraph*}
		} 
	\end{fmffile}\hspace{.2in}
%%%%%%%%%%%%%%%%%%%%%%%%%%%%%%%%%%%%%
	\begin{fmffile}{ctronem}
		\Scale[1.2]{
			\begin{fmfgraph*}(40,35)
				\fmfleft{i1,i2,i3,i4}
				\fmfright{o}
				\fmf{phantom,tension=5}{i1,v1}
				\fmf{plain,tension=-2}{i2,v1}
				\fmf{plain,tension=-2}{i3,v1}
				\fmf{phantom,tension=5}{i4,v1}
				\fmf{phantom,tension=5}{v2,o}
				\fmf{plain,left,tension=0.4}{v1,v2,v1}
				%\fmf{plain}{v1,v2}
				\fmfdot{v1}
				\fmf{phantom,label=\scalebox{.78}{${\;\; \z}$}}{o,v2}
				\fmflabel{\scalebox{.75}{${ \z_0}$}}{i2}
				\fmflabel{\scalebox{.75}{${ \z_0}$}}{i3}
			\end{fmfgraph*}
		}
	\end{fmffile}\;
	%%%%%%%%%%%%%%%%%%%%%%%%%%%%%%%%%%
	\begin{fmffile}{ctrtwom}
		\Scale[1.4]{
			%\begin{gathered}
			\begin{fmfgraph*}(30,30)
				\fmfleft{i1,i2,i3,i4}
				\fmfright{o}
				\fmf{phantom,tension=5}{i1,v1}
				\fmf{phantom,tension=-1}{i2,v1}
				\fmf{phantom,tension=-1}{i3,v1}
				\fmf{phantom,tension=5}{i4,v1}
				\fmf{phantom,tension=5}{v2,o}
				\fmf{plain,left,tension=0.4}{v1,v2,v1}
				%\fmf{plain}{v1,v2}
				\fmfdot{v1}
				\fmf{phantom,label=\scalebox{.65}{${ \z}$}}{o,v2}
			\end{fmfgraph*}
			%\end{gathered}
		}
	\end{fmffile}\;
	%%%%%%%%%%%%%%%%%%%%%%%%%%%%%%%%
	\begin{fmffile}{ctrresumm}
		\Scale[1.4]{
			\begin{fmfgraph*}(30,30)
				\fmfleft{i1,i2,i3,i4}
				\fmfright{o}
				\fmf{phantom,tension=5}{i1,v1}
				\fmf{phantom,tension=-1}{i2,v1}
				\fmf{phantom,tension=-1}{i3,v1}
				\fmf{phantom,tension=5}{i4,v1}
				\fmf{phantom,tension=5}{v2,o}
				\fmf{plain,left,tension=0.4}{v1,v2,v1}
				%\fmf{plain}{v1,v2}
				\fmfv{decor.shape=square,decor.size=2thick}{v1}
				\fmf{phantom,label=\scalebox{.65}{${ \z}$}}{o,v2}
			\end{fmfgraph*}	
		}	
	\end{fmffile}
	%	\end{center}
	\begin{center}
		\hspace{.15in}	(a)\hspace{.90in} (b) \hspace{.85in} (c) \hspace{.53in} \;(d) \hspace{.53in} (e)
	\end{center}		
	\caption{finite-temperature two-loop diagrams}
	\label{twoloop}
\end{figure}

The starting point of the OPT-improved thermal resummation can be taken to be the following renormalized action
\bea
S(\z) &=& -\int d^4x\; \fr12 \pa_\m \z \pa^\m \z  -\int d^4x\;\Big(\fr12 M^2 \z^2+ \fr{\l}{4!} \z^4\Big) -\int d^4x\;  \fr{3\n^4}{2\l}
\la{esabrenm} 
\eea
where
\be
M^2(T)\equiv \n^2+\fr{\l}{24} T^2.  \la{Massterm}
\ee
A word of caution: we take the expression $M^2(T)\equiv \n^2+\fr{\l}{24} T^2$ only for the purpose of {\em computing the loop contributions}: later when we sum up the classical and loop contributions (see eq. \rf{Mepq}), we use $M^2\equiv \n^2$ for the classical action so that the classical action is nothing but eq. \rf{flatsact}.\footnote{One may consider using the expression $M^2(T)\equiv \n^2+\fr{\l}{24} T^2$ for the classical action even when summing up the classical and loop contributions. This would be finite renormalization of the mass term. The qualitative conclusion on the CC problem remains unchanged. Here we follow the standard resummation and use eq. \rf{flatsact} for the classical action.} In the refined BFM one shifts the field according to \rf{srbfm}: 
\be
\z\ra \z+\tilde{\z} ,\quad  \tilde{\z}\equiv \z_c+\xi. \la{cs}
\ee 
With this shift one gets 
\bea
&&S(\z+\tilde{\z})= -\int d^4x\;\Big(\fr12 \pa_\m \tilde{\z} \pa^\m \tilde{\z}+\fr12 M^2 \tilde{\z}^2+\fr1{24} \l \m^{2\ve} \tilde{\z}^4\Big) -\int d^4x\;  \fr{3\n^4}{2\l \m^{2\ve}}\nn\\
&&-\int d^4x\; \Big[\fr12 \pa_\m \z \pa^\m \z +\fr12 M^2(\n,\tilde{\z}) \z^2\Big]-\int d^4x\;\fr{\l}{6}\m^{2\ve} \Big(\fr32 \tilde{\z}^2 \z^2+ \tilde{\z}\z^3+\fr14  \z^4\Big). \la{esaq} \nn\\
\eea
The one-loop effective potential is fairly standard and can be found in textbooks. Combined with the classical part it is given by 
\bea
V_{\mbox{classical+one-loop}}(\tilde{\z})
&=&\fr{3\n^4}{2\l} -\fr{\pi^2T^4}{90}-\fr{\Mt^4}{32\pi^2}\ln\fr{\bar{\m}e^{\g_E}}{4\pi T} +\fr{1}{24} \Mt^2 T^2\nn\\
&&\hspace{-.9in}+\fr12M^2\tilde{\z}^2-\fr{1}{12 \pi} \Mt^3T+\fr1{4!} \l \tilde{\z}^4+{\cal O}\Big(\fr{\Mt^6}{T^2}\Big). \la{1loopOPT}
\eea
At two-loop, things become significantly more involved. For the field-dependent part of the potential one can borrow the result obtained in \cite{Parwani:1991gq}\cite{Arnold:1992rz}. For our purpose it is also necessary to keep track of the field-independent terms, especially the temperature-dependent terms. Let us enumerate and categorize the two-loop-relevant diagrams. The first kind of the two-loop-relevant diagrams are those with two actual loops; they are given in Fig. \ref{twoloop} (a) and (b). The second kind are the diagrams involving one-loop counter vertices, Fig. \ref{twoloop} (c) - (e).
%%%%%%%%%%%%%%%%%%%%%%%%%%%%%%%%%%%%%%%%%%%%%%
%%%%%%%%%%%%%%%%%%%%%%%%%%%%%%%%%%%%%%%%
We illustrate the computation by taking the diagram in Fig. \ref{twoloop} (a). For Fig. \ref{twoloop} (a), including the pure temperature-dependent terms, one gets 
\bea
\m^{2\ve}V_{(a)} &=& \fr{1}{24}\l\m^{ 4\ve} \int d^4x\;< \z^4>\nn\\
&=&\fr{\l}{1152}T^4-\fr{\l}{192\pi}T^3 \Mt -\fr{\l}{768\pi^2}\Big(\fr1{\ve}+i_\ve +2\ln\fr{\bar{\m}}{ T}-2c_B-6\Big)T^2 \Mt^2\nn\\
&& +\fr{\l}4\fr{\Mt^3 T}{(4\pi)^3}\Big(\fr1{\ve}+2\ln\fr{\bar{\m}}{ T}-2c_B\Big)
	+\fr{\l}8\fr{\Mt^4}{(4\pi)^4}\Big(\fr1{\ve}+2\ln\fr{\bar{\m}}{ T}-2c_B\Big)^2
+{\cal O}(\l^{5/2} T^4) \nn\\
\eea
where
\be
\Mt^2(T,\tilde{\z})= \n^2 +  \fr{\l}{24} T^2   +\fr{\l}{2}  \tilde{\z}^2.
\la{Mmrel}
\ee
and 
\be
 i_\ve\equiv \ln\fr{\bar{\m}}{T^2}-4\ln3+c_H,\quad    c_H\equiv 5.3025,\quad  c_B\equiv \ln 4\pi -\g_E.
\ee
The result presented in \cite{Arnold:1992rz} is that it includes only the first line (without the first term since only the field-dependent terms were kept track of). The second line is not important for our purposes either: when analyzing the minima of the potential below, appropriate $\hbar$-scaling of the fields will be introduced, and given that $\hbar$-scaling, the terms in the second line are sub-leaking. The rest of the diagrams are as follows. 
\bea
\m^{2\ve}V_{(b)}&=&\fr{ \m^{2\ve}\l^2\tilde{\z}^2T^2}{48 (4\pi)^2}\Big(-\fr1{\ve}-i_\ve-\ln \fr{\bar{\m}^2}{T^2}-2\ln\fr{T^2}{\Mt^2}-2+c_H \Big)+\cdots\nn\\
\m^{2\ve}V_{(c)}
&=& \fr{1}{32(4\pi)^2}\l^2\m^{2\ve}T^2\tilde{\z}^2\; \Big(\fr1{\ve}+ i_\ve\Big)+\cdots  \nn\\
\m^{2\ve}V_{(d)}
&=&\fr{1}{48(4\pi)^2}M^2{\l} T^2\Big(\fr1{\ve}+ i_\ve\Big)+\cdots\nn\\
\m^{2\ve}V_{(e)}
&=&-\fr{\l}{48}\,T^2\,\Big[
\fr{T^2}{12}-\fr{\Mt T}{4\pi}-\fr{\Mt^2}{16 \pi^2}\Big(\fr1{\ve}+2\ln\fr{\bar{\m}}{ T}-2c_B\Big)+\cdots
\Big]. \la{eexp}
\eea
Fig. 1 (c) comes from the counter-term that removes the divergence in \rf{1lpp}. Again, there are $M$-dependent terms omitted here: given the $\hbar$-scaling, they are sub-leading. Combining the classical, one-loop, and all of the two-loop diagrams one can show after some algebra that
\bea
&&\hspace{.4in}V_{cl}+V_{1loop}+ \m^{2\ve} \Big(V_{(a)}+V_{(b)}+V_{(c)}+V_{(d)}+V_{(e)}\Big) \nn\\
&&\hspace{-.3in}=\fr{3\n^4}{2\l}  -\fr{\pi^2T^4}{90}
- \fr{\l}{1152}T^4   +\fr{ i_\ve}{48(4\pi)^2}{ M^2}{\l} T^2
+\fr{1}{24} \Mt^2T^2 +\fr12M^2\tilde{\z}^2 +\fr1{24} \l \tilde{\z}^4
\nn\\
&&\hspace{.3in}+\fr1{(4\pi)^2}\fr{1}{48 } \l^2\Big[\fr12i_\ve-\ln \fr{\bar{\m}^2}{T^2}-2\ln\fr{T^2}{\Mt^2}-2+c_H  \Big]\tilde{\z}^2T^2
\nn\\
&&\hspace{.1in}-\fr{1}{64\pi^2}\Big(\ln \fr{\bar{\m}^2}{T^2}-2c_B\Big) \,\Mt^4  -\fr{1}{12 \pi} \Mt^3 T -\fr{\l}{48(4\pi)^2}\Big(i_\ve -6\Big)T^2\Mt^2  
\la{Mepq}
\eea
where $M^2,\Mt^2$ are given in \rf{Massterm} and \rf{Mmrel}, respectively. Above, the $\fr1{\ve}$ terms have been removed by renormalization. Before implementing our OPT, we first find the minimum location and value of the potential. Instead of dealing directly with the $\tilde{\z}$ field, it is more convenient to treat $\Mt$ as the variable, from which the corresponding value of $\tilde{\z}$ can be easily read off. 
To better reveal the structure, we introduce the following rescalings\footnote{Since the upper limit of the imaginary time integration is taken as $\int^{\fr{\hbar}{T}}$, the temperature itself comes with an inverse power of $\hbar$. We keep this $\hbar$ implicit. (Equivalently one can introduce $\tilde{T}$ such that $T=\hbar\tilde{T}$, and use $\tilde{T}$ instead.) This simply means that the loop corrections remain small even with finite temperature, and the strength of the correction terms is determined by their overall $\hbar$ powers. 
}:
\bea
M^2 &\equiv& H^2 {\cal M}^2 =   \hbar {\cal M}^2 \nn\\
\tilde{M}^2 &\equiv& H^2 \tilde{{\cal M}}^2  = \hbar\tilde{{\cal M}}^2
\eea
where
\be
\hbar\equiv H^2
\ee
and display the $H$-dependence. For the terms that have explicit $\tilde{\z}$-dependence, we also make the following substitution
\be
\tilde{\z}^2=\fr{2}{\l}(\Mt^2-M^2).
\ee
This is of course to cast the potential into an expression having $\Mt$ as the variable.
Once the potential is rewritten in terms of the rescaled variables, it becomes clear why the previous omission of the higher-$M,\Mt$ terms is justified to the given order (which is two-loop). With these arrangements one can take the following form of the potential as the starting point of the analysis:
\bea
&&\hspace{.3in} V_{tot}=\frac{3 \n^4}{2 L}+H^2 \left(-\frac{{\cal M}^2 \n^2}{L}+\frac{\tilde{{\cal M}}^2 \n^2}{L}-\frac{1}{90} \pi ^2 T^4\right)\\
&&\hspace{-.5in}+H^4 \left(\frac{{\cal M}^4}{6 L}-\frac{{\cal M}^2 \tilde{{\cal M}}^2}{3 L}+\frac{\tilde{{\cal M}}^4}{6 L}+\frac{\tilde{{\cal M}}^2 T^2}{24}-\frac{L T^4}{1152}\right) -\frac{ H^5 \tilde{{\cal M}}^3 T}{12 \pi }
+{\cal O}(H^6).\nn
%%%%%%%%%%%%%%%%%%%%%%%%%%%%%%
\eea
The solution of $\fr{\pa V_{tot}}{\pa \tilde{{\cal M}}}=0$ gives the following minimum location in terms of $\tilde{{\cal M}}^2$:
\bea
\tilde{{\cal M}}^2&=&-\frac{3 \n^2}{H^2}-\frac{L T^2}{8}+{\cal M}^2+\frac{3 \sqrt{3} LT\sqrt{-  \n^2 }}{8 \pi } \nn\\
&&+ \frac{H^2}{128 \pi ^2 \n^2} \Big(  9  L^2 \n^2 T^2  +\sqrt{3} \pi\left(L T^2-8 {\cal M}^2\right) LT\sqrt{-  \n^2 }\Big)+\cdots.
\eea
Substituting this into $V_{tot}$ above yields the onshell potential. The PMS condition $\fr{\pa V_{tot} ( \n)}{\pa {\cal \n}}=0$ admits\footnote{The other branches of the solutions have undesirable features. For instance, in those branches the small $\tilde{{\cal M}}$-expansion is not justified.}
\be
\n=0
\ee
which implies
\bea
{\cal M}^2=\frac{1}{24}L T^2 
\eea
This translates into
\bea
\tilde{\z}^2 &=&-\frac{ T^2}{4}H^2+  i\frac{\sqrt{3L}  }{8 \pi }T^2 H^3 +\cdots
\eea
which then leads to the following value of the optimized potential:
\bea
V_{tot}&=&-\frac{\pi ^2}{90} \hbar  T^4+\frac{19  L }{1152}\hbar^2 T^4+{\cal O}\left(\hbar^{\fr52}\right).
\eea
This result confirms at two-loop the resolution of the CC problem proposed in the one-loop analysis in \cite{Park:2021ohu}.

\section{On potential QED asymptotic freedom}

In this section we discuss the possibility mentioned in \cite{Park:2018vci} in regards to potential asymptotic freedom of finite-temperature QED. We consider an Einstein-Maxwell system and revisit the issue by taking finite-temperature effects into account. Full analysis of the issue will require work solely dedicated to this endeavor. Postponing such full-scale investigation to future work, here we outline the  analysis necessary to get to the bottom of the problem. It turns out that to properly investigate QED asymptotic freedom at zero- or finite- temperature, one must undertake the whole renormalization procedure of the theory, not just renormalization of the CC. As for the case of CC renormalization, the finite part after divergence subtraction will be important. The conclusion that we draw based on the analysis below is that QED asymptotic freedom remains a reasonable possibility.

Inspired by the work of \cite{Robinson:2005fj} which studies the quantum gravitational effects on a Yang-Mills gauge coupling constant, QED asymptotic freedom was put forth in \cite{Toms:2010vy} wherein an additional term to the beta function was obtained. Let us quote eq. (12) of \cite{Toms:2010vy} for convenience:
\be
\b(E,e)=\fr{e^3}{12 \pi^2}-\fr{\k^2}{32\pi^2}\Big(E^2+\fr32 \L\Big)e. \la{Tresult}
\ee
The additional term, $-\fr{\k^2}{32\pi^2}E^2 e$, where $E$ denotes the characteristic energy scale, has the same form as the second term in the parentheses. This then led the author to propose potential asymptotic freedom in QED. The proposal of \cite{Toms:2010vy} as well as that of \cite{Robinson:2005fj} was debated in \cite{Pietrykowski:2006xy} and \cite{Ellis:2011}. In particular, it was suggested in \cite{Pietrykowski:2006xy} that such an additional term will be absent in de Donder gauge. As observed in those works the analysis must entail careful sorting-out of the tricky issue of gauge choice-(in)dependence of the effective action. This suggests the possibility that the potential gauge choice-dependence may be responsible for the different outcomes. It was noted in \cite{Park:2018vci} that finite-temperature effects should not be subject to such a gauge choice issue, and may lead to a term analogous to the one obtained in \cite{Toms:2010vy}. This must be so at least qualitatively: the fact that the presence of temperature makes contributions that appear inside the parentheses in \rf{Tresult} should be independent of the gauge choice.

 \begin{figure}
 	\hspace{-.3in}
 	\centerline{
 		\begin{minipage}[b]{7cm}
 			\epsfxsize=8cm
 			\epsfbox{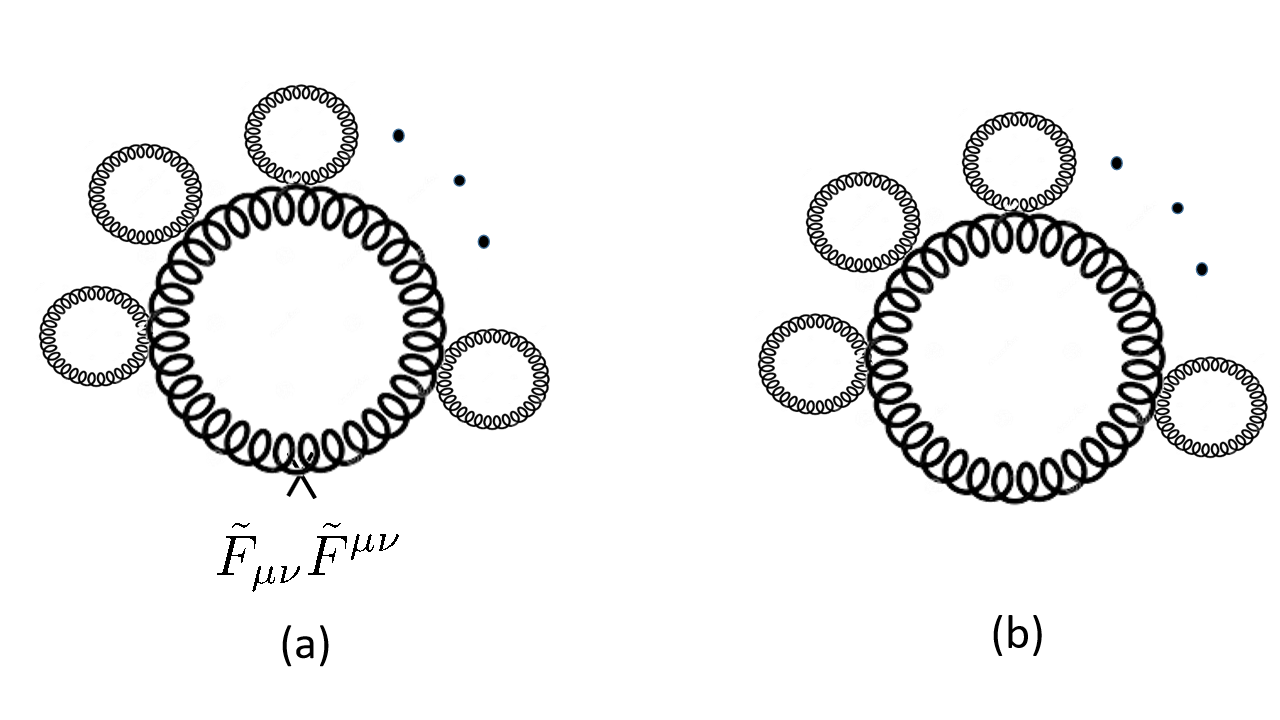}
 		\end{minipage}
 	}
 	\caption{Diagrams relevant for (a) gauge coupling  (b) cosmological constant; the curly lines represent gravitons}
 	\label{gfsquare}
 \end{figure}

The system considered in \cite{Robinson:2005fj} was a non-Abelian gauge theory coupled to gravity. A non-Abelian case has more diagrams than an Abelian case: for instance the graph in Fig. 1 of \cite{Robinson:2005fj} - which is present due to cubic gauge coupling - does not arise in an Einstein-Maxwell case. Once one considers finite temperature and resummation, there are diagrams  that additionally contribute to the beta function both in the Abelian and non-Abelian cases. These diagrams are of the type shown in Fig. \ref{gfsquare} (a). The explicit expression for the vertex represented by a cross in Fig. \ref{gfsquare} (a) can be found by examining the matter part of the Einstein-Maxwell action,
\bea
S_{matter} =-\fr14\int \sqrt{-\gh}\; \Fh_{\m\n}^2 . \la{EM2}
\eea
By introducing the fluctuation fields, $(h_{\m\n}, a_\m)$, and background fields, $(\tilde{g}_{\m\n},\At_\m)$,
\bea
\gh_{\m\n}\equiv  h_{\m\n}+\tilde{g}_{\m\n}\quad,\quad
\Ah_\m \equiv a_\m+\At_\m
  \la{gshift}
\eea
and expanding the matter action, one gets
\bea
&&\hspace{-.3in}S_{matter} =\int-\fr14 \sqrt{-\gt}\Big[\gt^{\m\n}\gt^{\r\s}-\gt^{\m\n}h^{\r\s} -\gt^{\r\s}h^{\m\n}+\fr12 \gt^{\m\n}\gt^{\r\s}h
+\gt^{\m\n}h^{\r\k}h_\k^{\s}  +\gt^{\r\s}h^{\m\k}h_\k^{\n}\nn\\
&&\hspace{-.7in}  -\fr12 \gt^{\m\n}hh^{\r\s}  -\fr12 \gt^{\r\s}hh^{\m\n}+h^{\m\n}h^{\r\s}  
+\fr18 \gt^{\m\n} \gt^{\r\s}(h^2-2h_{\k_1\k_2}h^{\k_1\k_2} )
\Big] \Big( f_{\m\r}f_{\n\s} {+}  2f_{\m\r}\Ft_{\n\s}+\Ft_{\m\r}\Ft_{\n\s} \Big) \nn\\
\eea
where $f_{\m\n},\Ft_{\m\n}$ denote the field strenth assocaited with $a_\r,\At_\r$, respectively.
The aforementioned vertex - which we call $V_{\tilde{F}\tilde{F}}$ - is given by the term containing $\Ft_{\m\r}\Ft_{\n\s}$ above:
\bea
&&\hspace{-.3in}V_{\tilde{F}\tilde{F}}\equiv -\fr14 \sqrt{-\gt}\Big[\gt^{\m\n}\gt^{\r\s}-\gt^{\m\n}h^{\r\s} -\gt^{\r\s}h^{\m\n}+\fr12 \gt^{\m\n}\gt^{\r\s}h
+\gt^{\m\n}h^{\r\k}h_\k^{\s}  +\gt^{\r\s}h^{\m\k}h_\k^{\n}\nn\\
&&  -\fr12 \gt^{\m\n}hh^{\r\s}  -\fr12 \gt^{\r\s}hh^{\m\n}+h^{\m\n}h^{\r\s}  
+\fr18 \gt^{\m\n} \gt^{\r\s}(h^2-2h_{\k_1\k_2}h^{\k_1\k_2} )
\Big] \Ft_{\m\r}\Ft_{\n\s}.  \nn\\
\eea
For renormalization of the gauge coupling, we consider $V_{\tilde{F}\tilde{F}}$ as an interaction vertex. (This is technically simpler than including $V_{\tilde{F}\tilde{F}}$ as part of the graviton kinetic term.) At zero temperature the diagram vanishes in dimensional regularization when the $\L$-CC term is either absent or not treated as a formal graviton mass term. This is not the case in the finite-temperature case. One should also consider Fig. \ref{gfsquare} (b) - which contributes to the CC. In other words the diagram in Fig. 2 (b) generates a CC term whose contribution to the graviton mass term must be taken into account when conducting the standard resummation.

Carrying out explicit evaluation of the diagrams and renormalization procedure will be very technically involved. For instance, computations of the diagrams in Fig. \ref{gfsquare} will require the terms in quartic order of the metric fluctuation $h_{\m\n}$ in the expanded Einstein-Hilbert action; this is quite lengthy. (It is given, e.g., in eq. (A.6) of \cite{Goroff:1985th}.) It is still possible to make some observations on the outcome, which connects us to the conclusion drawn in the beginning. The contribution of Fig. \ref{gfsquare} (b) to the CC will contain several types of terms, some of which will depend on the temperature. (Because of this, QED asymptotic freedom will occur conditionally, depending on the temperature.) Those terms will in turn contribute to the right-hand side of the beta-function calculation. As for the relative signs of those terms, we expect both signs to be present, which is different from the status of the $E^2$ term in \rf{Tresult}. The renormalization conditions will also matter since it will determine the finite parts. All of these seem to suggest that a systematic procedure of the renormalization, including the Newton's constant, would be necessary. One would also presumably need experimentally-derived inputs. In spite of these complications, the possibility of the QED asymptotic freedom seems reasonable. 

Lastly, recall that we have considered the presence of temperature for the reason stated before: finite-temperature effects should not be subject to the gauge choice-dependence issue. However, even if one considers the zero-temperature case, one would get various contributions, including the finite part, to the right-hand side of the beta-function calculation. This should conditionally imply QED asymptotic freedom.

\section{Conclusion}

It has recently turned out that quantization of gravity - which itself has been among the most evasive problems - holds the key to some other longstanding problems in theoretical physics. In this work we have carried out three exercises in the course of furthering progress. As established in the recent sequels, one of the essential ingredients for the FBQ is reduction of the support of the physical states. Whereas the PSC was derived in the ADM formalism in the earlier works, in the present work we have derived it in the 4D-covariant framework. Applying the quantized gravity setup to another longstanding CC problem, we have explored the finite-temperature effects in that context. Obviously, the crucial question is whether or not, in the case of a low temperature, the finite-temperature effects can be dismissed as unimportant and/or irrelevant. A qualitative scaling argument suggests that they should be crucial. It is shown that there exists a quantitative rationale - a variant OPT - that confirms the essential role of the finite-temperature effects in avoiding the CC fine-tuning. In another direction we have reexamined the possibility of QED asymptotic freedom put forth in \cite{Toms:2010vy}. We conclude that the QED asymptotic freedom remains a reasonable possibility, and to settle the matter it is necessary to conduct the entire renormalization procedure by paying close attention to the finite parts after divergence subtractions. 
%%%%%%%%%%%%%%%%%%%%%%%

In the body it was seen that the low energy sector is important for vacuum energy. Another qualitative way of seeing this is to examine the partition function in the canonical formalism, 
\be
Z=\sum_n <n|e^{-\fr{H}{T}}|n>.
\ee
A schematic notation is used above: the sum represents a combination of discrete sum and continuous momentum integration. Since the ultraviolet structure is determined by the momenta going to infinity, the structure will not be sensitive to the finite temperature. What is also clear from the expression is that the low energy states, i.e., states or virtual particles having energy $\sim T$ or lower, should be important. Put differently, the presence of the nonzero temperature changes the infrared structure of the theory. 
%%%%%%%%%%%%%%%%%%%%%%%%%%%%%%

Let us comment on an intriguing implication of the present vacuum energy analysis. Although temperature enters the analysis in a rather `mechanical' way, the fact that the CC is accounted for by a finite-temperature effect seems to reflect something profound about the nature of the spacetime. As well understood in cosmology, temperature is not just an indicator of the average kinetic energy of the particles (in the non-relativistic limit). It is also closely tied with expansion of the Universe. The present analysis reveals that it additionally serves as a measure of vacuum energy. Part of the vacuum energy comes from matter loops, and the vacuum energy in turn causes the expansion of the Universe. This inter-correlation of matter, vacuum energy, temperature, and expansion should be a manifestation of what may perhaps be described as the `organic' nature of the spacetime.

%\subsection{future directions}

\vspace{.3in}
There are several future directions:
\vspace{.1in}

One of the more urgent problems to better understand is in regard to boundary terms and conditions, regardless of the fact that much effort has been invested. For instance, the Neumann boundary condition that we have focused on results by not adding the YGH term \cite{Krishnan:2016tqj}. Will there be more general types of the Neumann boundary conditions? More narrowly, one considers a gauge parameter $\e^\m$ with the property $\e^\m \ra 0$ as $x^3\ra \infty$ in the Dirichlet boundary condition; will this restriction have to be lifted in the Neumann boundary condition? If so, how can such a transformation be distinguished from an LGT? Presumably such a restriction should be kept in the Neumann boundary condition as well. For one thing, lifting the restriction would interfere with partial integrations. This status of matter means that the boundary conditions are determined solely by added boundary terms. Further investigation is desirable for a more thorough understanding of boundary terms and conditions in general.

Consideration of finite-temperature effects is a crucial component in describing the thermal history of our Universe. Since the Universe was at higher temperatures in previous eras, it will be a meaningful endeavor to explore whether one could come up with a streamlining description covering the entire temperature range, say, from the electroweak era to the present. (The present results seem to signal toward an affirmative answer.) One must ultimately deal with the finite-temperature infrared problem. It will also be of great interest to examine whether or not the variant OPT could shed some new light on the possibility of first-order and second-order phase transitions.

\newpage
\appendix

\renewcommand{\theequation}{A.\arabic{equation}}
\setcounter{equation}{0}

\section{Glossary of some terms}

{\bf FBQ} (foliation-based quantization): It is an approach of quantizing gravity, proposed in \cite{Park:2014tia}. It has ingredients of both canonical quantization and covariant quantization. One of the key features is that appropriate and complete gauge-fixing leads to the physical sector of the theory, which has support on a hypersurface located in an asymptotic boundary.\\ 

\ni{\bf PSC} (physical state condition): Just as one finds two polarization components after fixing the gauge symmetry of Maxwell's theory, one can fix the diffeomorphism symmetry of gravity and obtain the physical spectrum of the theory. The procedure is more involved due to the large amount of the gauge degrees of freedom. At a more technical level, the difficulty of the procedure is attributed to the fact that first-class constraints are involved. The PCS was initially obtained in Arnowitt-Deser-Misner formalism. When solving the first class constraint, the commutation relation between Lie and covariant derivatives \cite{Kobayashi} plays a crucial role.\\

\ni{\bf BFM} (background field method): The background field method is a convenient way of computing the 1PI effective action by splitting the fields into fluctuations and backgrounds, followed by integrating out the fluctuations. The refined version employed in the present work (and its sequels) is based on eq. (16.1.17) of \cite{Weinberg2}, wherein it was called the external field method. When applied to a metric field, care must be exercised in dealing with the trace piece of the fluctuation. More pedagogical technical details of the present method were given in \cite{Park:2014noa} and \cite{Park:2015ota}. More thorough applications and some subtleties of the method can be found in \cite{Park:2018vci}.\\

\ni{\bf OPT} (optimal perturbation theory): Optimal perturbation theory has its origin in the variational principle in quantum mechanics. Not surprisingly, some of the early applications were in the context of quantum mechanics, such as an anharmonic oscillator. In the work of \cite{Stevenson:1981vj}, the notion was applied to renormalization condition of quantum field theory. The resulting perturbation theory was named OPT.\\

\ni{\bf PMS} (principle of minimal sensitivity): The values of the variational parameters in quantum mechanics are determined in such a way as to minimize the energy of the wave function. By invoking renormalization group invariance of physical quantities, it was put forth in \cite{Stevenson:1981vj} that an optimal renormalization condition must be one that is stable under its variation. The variational principle in this specific context of the renormalization conditions was dubbed the principle of minimal sensitivity.\\

\newpage
%%%%%%%%%%%%%%%%%%%%%%%%%%%%%%%%%%%%%%%%%%%%%%%%%%%%%%%%%%%%%%%%

\end{document}